\documentclass[12pt]{article} 

\usepackage{times, url} 
\usepackage{indentfirst}
\usepackage{graphics,color}
\usepackage{geometry}
\usepackage{fancyhdr}
\begin{document} \title{Energy
Emission by 
Quantum Systems in an Expanding FRW Metric}
\author{D.P. Sheehan\\Department of Physics, University of San Diego\\5998 Alcala
Park, San Diego, CA 92110\\dsheehan@sandiego.edu\\
V.G. Kriss\\Physics Department, Lewis-Clark State
College\\Lewiston, ID 83501} \maketitle

\begin{abstract} 

Bound quantum mechanical systems not expanding with the comoving frame of
an expanding, flat FRW metric are found to release energy at
a rate linearly proportional to the local Hubble constant ($H_{o}$) and
the systems' binding energy ($E_{b}$); {\em i.e.}, $\dot{E} = H_{o}
E_{b}$.
Three exemplary quantum systems are examined.  For systems with early
cosmological condensation times | notably hadrons | time-integrated energy
release could have been significant and could account for an appreciable fraction of the dark matter inventory.\\
  PACS: 95.35.+d, 98.80.-k, 98.80.Cq

Key Words: cosmology, metric expansion, dark matter, dark energy, energy
conservation.\\

\end{abstract}

\newpage

\section{Introduction} 

A perennial unresolved issue intrinsic to the standard cosmological model
is the scale length at which the expansion of the spacetime metric ceases. 
Concerns about this date back at least to McVittie [1], Einstein and
Straus [2] and Noerdlinger and Petrosian [3].  These have been taken up
with increasing earnest in recent years by Anderson [4], Bonner [5],
Cooperstock, et al. [6], and others. Recent evidence for dark energy,
which accelerates cosmic expansion and which presumably permeates the
vacuum at all scale lengths, makes the metric expansion cut-off
increasingly salient [7-9]. 

Most recent studies [4-6] have concentrated on issues of particle dynamics
in the presence of cosmological expansion.  These studies have found that
multiparticle systems at size scales below the scale of the perceived
Hubble flow should not appreciably expand along with the comoving metric. 
However, a facet of this problem that has not received adequate attention
is the relationship of cosmic expansion to the behavior of quantum
systems, especially with regard to energy radiation. 

This paper explores the physical ramifications of extending the principle
of cosmological metric expansion down to quantum mechanical scale lengths. 
We find that negative-energy, bound quantum systems which collapse in the
comoving frame against the expanding cosmological metric should release
energy at a rate ($\frac{d E}{d t} \equiv \dot{E}$) that is linearly
proportional to the
local Hubble constant ($H_{o}$)  and to the system binding energy
($E_{b}$); that is:  $\dot{E} \sim H_{o} E_{b}$.  Exemplary
quantum systems are shown to follow this prescription. For quantum systems
that condensed early out of the primordial fireball | particularly hadrons
| the time-integrated contraction energy since condensation might have
been sizable.  In the present era, the Hubble constant is sufficiently
small to make this energy release rate negligible.  Hereafter, we will
refer to this type of energy as {\em metric contraction energy} (MCE).

In this paper, a {\em system} will refer to any multiparticle assemblage,
for instance, baryons, nuclei, atoms, molecules, planets, stars, or
galaxies.  We assume the Friedmann-Robertson-Walker line element, $ds$,
for a spatially-flat, open universe:  $ds^{2} = dt^{2} + dl^{2} = dt^{2} -
a^{2}(t)[ dx^{2} + dy^{2} + dz^{2} ]$, where $t$ is time, (x,y,z) are the
spatial coordinates, and $a(t)$ is the metric expansion parameter, which
defines the Hubble parameter via $H \equiv \frac{\dot{a(t)}}{a(t)}$. 

The remainder of this paper is organized as follows.  Primary assumptions for MCE
emission are introduced and three archetypical quantum systems are considered for MCE
emission.  General expressions for instantaneous and time-integrated MCE emission are
derived.  The latter is applied to quantum systems that condensed from the primordial
fireball.  Finally, issues of global energy conservation are raised.\\

\section{Model Assumptions}
MCE emission follows directly from the following assumptions:
\begin{quote} 

A) Metric expansion proceeds at all gravitationally classical scale
lengths, meaning scale lengths greater than the Planck length ($L_{p} =
\sqrt{\frac{h G}{c^{3}}} \simeq 10^{-35}$m).

B) Bound systems contract in the comoving frame in such a way that they
appear spatially unaffected in the proper frame.

C) Energy emission is possible in quantum systems since they undergo
transitions diferently than classical systems. 

\end{quote}
We examine each of these assumptions for physical reasonableness.

{\em Assumption A:} It is normally assumed that cosmic metric expansion
ceases at scale lengths below those of Hubble flow.  This view has been
championed by many, but this does not settle the issue.  In view of the
evidence from the supernova searches [7,8] and WMAP results [9] that
suggest the universe is filled with an ubiquitous negative-pressure dark
energy, expansion should proceed down to the smallest classically allowed
gravitational scale length ($L_{p}$).  This would be at scale lengths 20
orders of magnitude smaller than the proton, roughly 60 orders of
magnitude smaller than where Hubble flow is typically observed, thereby
subsuming all known physical systems. Even theoretically, this is not so
untenable a position as it may seem since the general relativistic field
equations, from which cosmological expansion is derived, presumably hold
sway at all such gravitationally classical scales.  The field equations
contain the cosmological constant (as the simplest explanation for dark
energy) which appears in the vacuum energy density that is the source for
the long-range gravitational field in effective quantum field theories.

Metric expansion | as inferred from the non-expansion of material systems | is not observed
to proceed in regions of space in which the mass-energy density exceeds the critical
density, $\rho_{critical} \simeq \frac{H_{o}^{2}}{G}$, where $H_{o}$ is the local Hubble
constant.  This is equivalent, however, to having metric expansion in fact proceed
identically on all gravitationally classical scale lengths, but having it not proceed if the
local self-gravitational acceleration ($a_{g}$ radially inward) exceeds the Hubble
acceleration ($a_{H}$ radially outward); that is, $a_{g} > a_{H} \sim r H_{o}^{2}$, where r
is the system scale length. (This criterion for non-expansion also applies to forces other
than gravity.)  This interpretation reproduces local physical behavior while remaining
faithful to the strict FRW formalism.  In summary, Assumption A | that
metric expansion proceeds at
all
gravitationally classical scale lengths | is not contradicted by available
physical evidence, and it is consistent with the standard general relativistic
interpretation of the metric.\\

{\em Assumption B:} That bound systems contract in the comoving frame to
counter the expansion of space follows naturally from assumption (A) and
is consistent with observational evidence such as the recent
investigations of the value of the fine structure constant in distant
galaxies [10].  The contraction of bound systems (such as hydrogen atoms)
proceeds in such a way that fundamental energy levels of quantum systms
are unchanged over measurable time-averaged periods.  Thus, the energy
structure of the systems appears unaffected, in agreement with
observation.\\

{\em Assumption C:} It is through the fundamental differences by which
classical and quantum systems undergo energy state transitions that we
propose the MCE originates.  A classical system can spatially contract
continually and continuously against the comoving metric expansion and,
thereby, maintain a fixed size ({\em i.e.}, shrink within an expanding
comoving frame).  Furthermore, its inertial response (follow-through)
favors its continuous local contraction which counteracts metric
expansion.  As a result, we predict little or no continuous MCE emission
from classical systems [11].

In contrast to classical systems, quantum systems radiate energy
intermittantly and discontinuously, and they do not exhibit inertial
follow-through.  As a result, unlike classical systems, we posit that they
can follow the expansion of the comoving metric during time intervals
between radiating (Assumption A).  If they expand along with the comoving
metric between emission, but intermittantly collapse back to their
original fixed proper sizes, then they should emit MCE, analogously to
classical systems ({\em e.g.}, gas clouds) and more typical quantum
systems ({\em e.g.}, hydrogen atoms).  We emphasize
that quantum MCE emission is not due to change in quantum number, but is
due to change in spatial scale only. Emission should continue as long as
they are embedded in an expanding spacetime metric, but emission rate
clearly depends on the metric expansion rate [12]. Assumption C is
consistent with the standard physical interpretation of energy release by
classical and quantum systems.  It presumes, of course, the validity of
Assumptions A and B.\\ 

Justification for the emission of radiation is provided by the
interpretation
of recent evidence for accelerated cosmic expansion and dark energy
[7-9].  If every point in space is a locus of expansion and if expansion
is indeed driven by intrinic properties of the vacuum, then it is
reasonable to assume that all interactions and bound systems described by
Feynman diagrams | even at the tree level | should be modified (albeit
slightly) by the expansion of space as the "moving" vacuum passes through
the system.  This will be true in both the comoving and proper frames.  In
the comoving frame, particles in a shrinking bound system will slide
through a vacuum at rest.  On the other hand, in the bound system's proper
frame, a moving vacuum will slide past the bound particles. The exact
mechanism producing the radiation remains obscure since we do not have a
good theoretical model for describing how expansion (via dark energy?)
affects the vacuum.  There is a suggestion that the energy producing
mechanism has similarity to the Unruh effect, where spatial expansion
plays a role similar to acceleration in providing an energy source.  A
better model for the mechanism of spatial expansion is required before we
can say more on this subject.

\section{MCE in Quantum Systems} \noindent {\bf 1-D Systems}\\ In this section, several
quantum systems are examined for MCE emission, subject to Assumptions A-C above.  To begin,
consider an archetypical quantum mechanical wave comoving in an expanding R-W metric, as
described by Peebles [14].  (This model is commonly used to illustrate the observed
redshifting of radiation from distant objects in an expanding universe.  We adopt Peebles'
notation.)  Let a normal mode wave form a one-dimensional closed circular loop (radius
$a(t)$); that is, an integer number of wavelengths ($n$) span the loop circumference (Fig.
1).  The loop is of cosmic proportion.  The time evolution of the wavelength ($\lambda (t)$)
follows the radial metric expansion parameter: $\lambda (t) \propto a(t)$.  Since the
expansion is adiabatic, there are no discontinuous changes in state or occupation numbers; 
thus the wavelength continuously increases, producing a redshift, and the mode energy
continuously decreases. 

If, from time to time, this mode were to decouple from the comoving metric and
collapse down to a smaller, fixed, fiduciary radius $a(t_{o})$, then an energy
change is expected.  For a photon ($E = \frac{h c}{\lambda} = \frac{n
\hbar c}{a(t)}$), the photonic energy gain resulting from its intermittant contraction from
$a(t)$ to $a(t_{o}) \equiv a_{o}$ (with $t \geq t_{o}$) would be: 
\begin{equation}
\Delta E = n \hbar c \left[ \frac{1}{a_{o}} - \frac{1}{a(t)}
\right]
\end{equation}

Returning to Fig. 1, now let the loop be opened, the wave straightened and placed
between two endpoints. Metric expansion now corresponds to the separation of the
endpoints.  Wavelength can be written:  $\frac{n \lambda}{2} = a(t)$; with ($n =
1,2,3 ...$).  If one applies the deBroglie relation for wavelength and energy, one
obtains the bound state energies for a particle in a one-dimensional square well
(with comoving walls): 
\begin{equation}
E_{n}(t) = \frac{ n^{2} h^{2}}{8 m a^{2}(t)}
\end{equation}

\noindent Thus, the redshift of radiation in an expanding spacetime metric can be
seamlessly translated into the one-dimensional particle-in-box with expanding
walls. 

To estimate the MCE emission rate for this system, let the system expand with the comoving
frame for time $\delta t$, afterwhich it undergoes a transition back to its original size. 
(Again, we emphasize that this transition {\em does not} involve a change in quantum number,
but a change in spatial scale only.)  Via Hubble's law, one can write the metric expansion
as \begin{equation} a(t_{o} + \delta t) = a_{o} + v \delta t = a_{o} + H_{o} {\cal L} \delta
t, \end{equation}

\noindent where ${\cal L}$ is the scale size over which expansion occurs.  The
energy of the comoving, expanding ground state ($n=1$) becomes: 
\begin{equation}
E_{1}(\delta t) = \frac{h^{2}}{8 m [a_{o} + H_{o} {\cal L} \delta t]^{2}} =
\frac{h^{2}}{8 m a_{o}^{2} [1 + \frac{H_{o} {\cal L} \delta t}{a_{o}}]^{2}}.
\end{equation}

\noindent Under the assumption that
the system
expansion during time interval
$\delta t$ is
small compared with ${\cal L}$ or $a_{o}$, this can be 
expanded to first order ($ \frac{H_{o} {\cal L} \delta t}{a_{o}} \ll 1$),
rendering:
\begin{equation}
E_{1}(\delta t) \simeq \frac{h^{2}}{8 m a_{o}^{2}} \left[ 1 - \frac{2 H_{o} {\cal
L}}{a_{o}} \delta t \right].
\end{equation}

\noindent Letting the proper physical distance ${\cal L}$ be the system scale size
${\cal L}
= \frac{a_{o}}{2}$, one has
\begin{equation}
E_{1}(\delta t) \simeq \frac{h^{2}}{8 m a_{o}^{2}} \left[ 1 - H_{o} \delta t
\right] = E_{1}(0) \left[ 1 - H_{o} \delta t \right],
\end{equation}

\noindent from which the MCE increment is $\delta E_{mce} \equiv
E_{1}(\delta t) - E_{1}(0) = - E_{1}(0) H_{o} \delta t$.  In the continuum
limit ($\delta t \longrightarrow dt$ and $\delta E_{mce} \longrightarrow
dE_{mce}$), recognizing that $E_{1}(0) = E_{b}$ is the system's binding
energy, the MCE emission rate from the ground state particle-in-box can be
written [15]

\begin{equation}
\dot{E}_{mce} = - H_{o} E_{b}.
\end{equation}

\noindent These first two examples are positive-energy bound states so the
energy change for the universe (in the comoving frame) is negative
($\dot{E}_{mce} < 0$);  however, were these negative-energy
states, typical of realistic bound systems, then $\dot{E}_{mce} >
0$. This latter result suggests that negative-energy bound quantum states
should radiate MCE at a time-averaged rate which is linearly proportional
to their bound state energies and the present epoch Hubble constant,
$H_{o}$.  MCE represents {\em radiated} energy rather than simply an
energy {\em shift} such as the Lamb shift.  Given the low value of $H_{o}$
in the present era ($H_{o} \sim 10^{-18}$ sec$^{-1}$), the MCE emission
rate can be shown to be small for standard bound systems;  however, it
will be shown that, for earlier epochs, $\dot{E}_{mce}$ could
have been considerable.\\

\noindent {\bf Hydrogen Atom}\\ A more realistic case of cosmic expansion
affecting a bound state is that of the hydrogen atom.  This system was
investigated for expansion by Bonner [5] who studied the ground state Bohr
atom in a flat FRW metric. He concluded that cosmological expansion had
negligible effect on the size of the atom.  We, too, presume that the
long-time average size and energy of the atom are essentially constant,
but that these are due to the atom's intermittant quantum mechanical
contractions against the comoving frame. We take as the size of the ground
state hydrogen atom the Bohr radius, $b$. Unlike the case of the 1-D
square well, with size $a_{o}$, there is no specific expansion parameter
for the hydrogen atom, so we will take a different route to an expression
for MCE: time-dependent perturbation theory.  This proceeds from the
relations: $i \hbar \frac{\partial | \psi \rangle}{\partial t} = {\cal H}
| \psi \rangle$
and ${\cal H}(t) = {\cal H}^{o} + {\cal H}^{1}(t) $, where ${\cal
H}^{1}(t)$ is the first-order perturbation.

Since the perturbation arising from metric expansion is small and is slowly
varying compared with the system transitions [16], we can adopt the adiabatic
approximation; thus, we can consider the perturbation as being time-independent
over the interval $\delta t$ and time-independent perturbation theory can be
validly employed.  For the ground state hydrogen atom, we write
\begin{equation}
{\cal H} \psi = - \frac{\hbar^{2}}{2 m} \frac{\partial^{2}}{\partial
r^{2}}
\psi (r^{1}) - \frac{e^{2}}{4 \pi \epsilon_{o}} \frac{1}{r^{1}} \psi (r^{1}) = E
\psi (r^{1})
\end{equation}

\noindent Inserting the Hubble-expanded radius, $r^{1}(\delta t) = r + H_{o} {\cal
L} \delta t$, the potential energy term becomes
\begin{equation}
V(r^{1}) = - \frac{e^{2}}{4 \pi \epsilon_{o} [r + H_{o} {\cal L} \delta t]} \simeq 
 - \frac{e^{2}}{4 \pi \epsilon_{o} r} \left[ 1 - \frac{H_{o} {\cal L} \delta t}{r}
\right].
\end{equation}

\noindent From this, the perturbation to the Hamiltonian for the hydrogen ground
state is inferred to be:
\begin{equation}
H^{1}(r) = \frac{e^{2}}{4 \pi \epsilon_{o} r^{2}} (H_{o} {\cal L} \delta t).
\end{equation}

The first-order energy correction to the unperturbed wavefunction $\psi^{o}_{n}$ is $\delta
E^{1} = \langle
\psi^{o}_{n} | {\cal H}^{1} | \psi^{o}_{n} \rangle$.  For the unperturbed ground
state hydrogen wavefunction, given by $\psi^{o}_{1} (r) = \frac{1}{\sqrt{\pi
b^{3}}} exp[- r/b]$, evaluating Eq.(10) and noting $\langle r^{-2} \rangle =
\frac{2}{b^{2}}$, one obtains
\begin{equation}
\frac{\delta E^{1}}{\delta t}  = \frac{e^{2}}{4 \pi \epsilon_{o}} H_{o} {\cal L}
\left( \frac{2}{b^{2}} \right).
\end{equation}

If one now takes the proper distance ${\cal L}$ to be ${\cal L} = \langle
\psi^{o}_{1} | r | \psi^{o}_{1} \rangle = \langle r \rangle = \frac{b}{2}$, then in the
continuum limit, one obtains for the first-order MCE emission rate from the ground
state hydrogen
atom
\begin{equation}
\dot{E}_{mce} = 2 H_{o} E_{b}.
\end{equation}

\noindent Within a factor of order unity, this agrees with the principal result of the
square well.\\

\noindent {\bf Homogeneous Potentials}\\ By invoking the virial theorem, these
results can be generalized for arbitrary homogeneous, negative-energy potentials
of the form $V(r) = \kappa r^{n}$.  The total system energy is $E = \bar{T} +
\bar{U}$, where $\bar{T}$ and $\bar{U}$ are the time-averaged kinetic and
potential energies, and $E = (\frac{n}{2} + 1) \bar{U}$.  The metric-perturbed
energy expectation value is
\begin{equation}
\langle E^{1} \rangle \equiv \langle \psi | {\cal H}^{1} | \psi \rangle = \langle
\psi |
(\frac{n}{2} + 1)[\kappa(r +
H_{o}
{\cal L} \delta t]^{n} | \psi \rangle \simeq \kappa (\frac{n}{2} + 1) \langle \psi | r^{n}
[1 + \frac{H_{o} {\cal L} \delta t}{r}]^{n} | \psi \rangle,
\end{equation}

Taking $\frac{{\cal L}}{r} \equiv \chi$ to be a constant, then to first order Eq.
(13) becomes 
\begin{equation}
\langle E^{1} \rangle \simeq \kappa (\frac{n}{2} + 1) \left[ \langle \psi | r^{n}
| \psi
\rangle + n H_{o} \chi \delta t \langle \psi | r^{n} | \psi \rangle \right] = \langle
E_{b} \rangle + n \chi H_{o} \delta t \langle E_{b} \rangle
\end{equation}

The contraction energy release is $ \delta E \equiv \langle E^{1} \rangle -
\langle
E_{b} \rangle = n \chi H_{o} \langle E_{b} \rangle \delta t$.  In the continuum
limit, and with $n
\chi \sim 1$, one has for the time-averaged MCE release rate
\begin{equation}
\dot{E}_{mce} \simeq H_{o} \langle E_{b} \rangle,
\end{equation}

\noindent as was found for the previous explicit cases. \\

\noindent {\bf Time-Integrated MCE -- Cosmological Implications}\\ The adiabatic
approximation can be extended to longer times periods, for which $H_{o}$ is no
longer a constant, but varies with time.  In most expansion models, $H$ can be
written $H(t) = \frac{\alpha}{t}$, where $0 \leq \alpha \leq 1$ and $t$ is the age
of the universe.  For a radiation-dominated universe, $\alpha_{r} = \frac{1}{2}$,
while for a matter-dominated universe $\alpha_{m} = \frac{2}{3}$. The time of
cross-over from radiation to matter dominance, $t_{eq}$, is roughly $t_{eq} \simeq
10^{11}$ sec.  Here $\Omega$ is assumed to be the critical value, $\Omega = 1$.
(In this study, we do not consider the effects of accelerated expansion due to
dark energy [7-9], but, in principle, these can be incorporated into this
model.)
As the universe expands and cools, multiparticles systems (e.g.  baryons, nuclei,
atoms)  condense as the temperature falls below their rest masses and binding
energies.  From Eq. (15), the time-integrated MCE emission since the time of
system condensation, $t_{c}$, will be:
\begin{equation} 
E(t_{c} \rightarrow t) = \int^{t}_{t_{c}} \langle E_{b} \rangle \frac{\alpha}{t} dt =
\langle E_{b} \rangle \left[ \int^{t_{eq}}_{t_{c}} \frac{\alpha_{r}}{t} dt +
\int^{t}_{t_{eq}} \frac{\alpha_{m}}{t} dt \right]  =
\langle E_{b} \rangle \{ \frac{1}{2} ln(\frac{t_{eq}}{t_{c}}) + \frac{2}{3}
ln(\frac{t}{t_{eq}}) \},
\end{equation} 

\noindent with the proviso, $t_{eq} \geq t_{c}$.  

In Table I are given the approximate condensation times, binding energies, and net mass
gains for several multiparticle quantum systems of cosmological importance.  Condensation
times are taken from the standard cosmological model. The net mass gain, G, is the average
fractional mass-equivalent of energy released from the condensation time to the present era,
as calculated from Eq. (16).  As indicated in Table I, the dominant source of MCE derives
from hadrons, specifically baryons. Emissions from nuclei, atoms and molecules are orders of
magnitude less, both by virtue of their significantly lesser binding energies and also due
to their more recent condensation times.  For the present era ($t \sim 5 \times 10^{17}$
sec), the mass-energy of the baryonic MCE radiated since the hadron condensation should be
about 3 times the baryonic mass in the universe; if one includes estimates of additional MCE
generated near the quark-hadron condensation time due to thermal effects, the total
time-integrated MCE could have been as much as roughly 10 times the baryonic mass-energy
[13,17].

\section{Discussion}

The analysis above indicates that non-expanding, multiparticle systems in an expanding
universe should release energy (MCE) continuously.  In this study, we do not speculate as to
what form MCE takes or whether it survived the effects of subsequent metric expansion [18].
We
also make no account of the underlying source of MCE; we simply argue that it should be
radiated.  Clearly, it is tied to metric expansion, but we do not speculate as to whether
MCE draws from global expansion energy, dark energy, or some other energy reservoir. In this
model, the choice has been made to preserve the metric expansion for all gravitationally
classical scale lengths even at the possible expense of global energy conservation.  Since
the former seems demanded by the standard cosmological model while the latter does not, this
seems a defensible choice. 

Energy conservation on universal distance scales remains an open question.  It is well-known
that in the standard model the cosmological redshift of the cosmic background radiation does
not conserve energy globally.  To resolve this paradox, it is usually pointed out that the
general theory of relativity does not contain a general global energy conservation law.
Schemes have been proposed for mining energy from the expanding universe, not all of which
appear to satisfy global energy conservation.  Davies [19] has proposed a simple means by
which to mine energy from a deSitter universe, but concludes that this "second helping" is
not a 'free lunch.' Harrison [20], on the other hand, proposes a mining scheme for Friedmann
universes using a network of strings which he claims {\em is} a 'free lunch.' A consensus
has not
been reached as to the meaning of these gedanken experiments, however, it keeps at the fore
unsettling and unsettled issues about energy conservation. 

This MCE model bears strong resemblance to Harrison's tethered body gedanken experiment with
the Harrison tether replaced by force-mediating exchange particles.  If so, many of the same
issues Harrison raises concerning energy conservation might also apply here.  This ambiguity
seems tracable to the lack of global energy conservation in the standard cosmological model. 
It is unclear whether the ultimate source of the energy is found in cosmic expansion or
whether it is "nascent" energy, as argued by Harrison [20].  Additional analysis or suitable
experimental tests appear necessary to settle these issues. \\

\newpage

\begin{center}
{\bf References}
\end{center}
\begin{enumerate}

\item McVittie, G.C. Mon. Not. R. Astron. Soc. {\bf 93} 325 (1933).\\

\item Einstein, A. and Straus, E.G., Rev. Mod. Phys. {\bf 17} 120 (1945).\\

\item Noerdlinger, P.D. and Petrosian, V., Astrophys. J. {\bf 168} 1 (1971).\\

\item Anderson, J.L., Phys. Rev. Lett. {\bf 75} 3602 (1995).\\

\item Bonner, W.B., Class. Quant. Grav. {\bf 16} 1313 (1999); Mon. Not. R. Astron. Soc.
{\bf 282} 1467 (1996); Gen Rel. Grav. {\bf 32} 1005 (2000).\\

\item Cooperstock, F.I., Faraoni, V., and Vollick, D.N., Astrophys. J. {\bf 503}
61 (1998).\\

\item Perlmatter, S., et al., Astrophys. J. {\bf 517} 565 (1999).\\

\item Riess, A., et al., Astron. J. {\bf 116} 1009 (1998).\\

\item Turner, M.S., astro-ph/0207297v1 (2002); Bennet, C.L., et al.,
Astrophys. J., in press, (astro-ph0302207)  (2003).\\

\item Ashenfelder, T., Mathews, G.J., and Olive, K.A., astro-ph/0404257
(2004).

\item One may argue, however, that exceptions exist.  For instance, the release of
gravitational potential energy by a molecular cloud can be traced ultimately to
metric expansion of the gas during an earlier epoch of the universe and therefore
qualifies as classical (not quantum) MCE. This MCE release is neither continuous,
nor sizable compared with what quantum systems should display, nor is it the
result of contemporary metric expansion; rather, it is a relic energy release.\\

\item In order to plausibly qualify for collapse in the comoving frame, a
quantum system
should
expand during the time between transitions by at least the minimum gravitational
scale length: $L_{p}$.  Additionally, to respect causality, the time between
transitions should equal or exceed the light travel time across the system ({\em
i.e.}, $\tau_{L} = \frac{L_{QM}}{c}$).  Causality-challenging EPR effects aside,
this constraint is generally met by real systems.  For systems subject to Hubble
expansion, these two constraints can be shown to translate into a single condition
for MCE emission: $L_{QM} > \sqrt{L_{p} L_{u}}$.  Here $L_{u}$ is the visible
scale size of the universe.  If the time interval between transitions ($\tau_{t}$)
is longer than $\tau_{L}$ | as is the case for many quantum systems | then the
minimum $L_{QM}$ will decrease accordingly; the condition for MCE emission can
then be written:  $L_{QM} > \frac{L_{p}}{\tau_{t} H_{o}}$.\\

\item Sheehan, D.P., unpublished, (1995).\\

\item Peebles, P.J.I., \underline{Physical Cosmology} (Princeton Univ. 
Press, Princeton, 1993), pg. 97.\\

\item Analysis of the 3-D spherical well bound states returns the same results as
the 1-D case.\\

\item Under Assumption A, for an atom to `expand' with the metric but remain `fixed'
in size, requires intermittant contractions on a time scale short compared with
observational time scales.\\

\item This baryonic MCE mass gain roughly matches the currently estimated dark matter
inventory.\\

\item If the MCE were in the form of relativistic particles ({\em e.g.}, photons), subject
themselves to metric expansion, then little relic MCE would be expected to survive to the
present day.\\

\item P.C.W. Davies, Phys. Rev. D {\bf 30} 737 (1984).\\

\item E.R. Harrison, Astrophys. J. {\bf 63} 446 (1995).\\

\end{enumerate}

-\\

\noindent {\bf Acknowledgements:}  The authors thank Ms. K. Adkins and Mr. Matthew S. Dominick for technical assistance.  W.F. Sheehan is thanked for stimulating conversations.

\newpage

Table I: Sources of MCE.  Notation: $t_{c} \equiv$ condensation time after Big
Bang; $\frac{E_{b}}{mc^{2}} \equiv$
ratio of binding energy to rest mass energy; $C \equiv (\frac{1}{2}
ln(\frac{t_{eq}}{t_{c}}) + \frac{2}{3} \ln(\frac{t}{t_{eq}}))$; $G \equiv C \frac{E_{b}}{mc^{2}} \equiv$ average fractional gain in mass of composite system over the lifetime of the universe. ($t \simeq 5 \times 10^{17}$ sec, $t_{eq} \simeq 10^{11}$ sec).

\begin{table}[h]
\begin{center}

\begin{tabular}{| c | c | c | c | c | c |} \hline
\bf Particles & \bf System & \bf $\bf t_{ c}$(sec) & \bf $\bf \frac{E_{b}}{mc^{2}}$ & \bf C & \bf G\\  \hline \hline
\rm quark & \rm hadron & \rm $10^{-6}$ & \rm $10^{-1}$ & \rm 30 & \rm 3\\  \hline 
\rm nucleon & \rm nucleus & \rm $10^{2}$ & \rm 2x$10^{-3}$ & \rm 20 & \rm 0.04\\  \hline 
\rm $\rm e^{-}$/nuclei & \rm atom & \rm $10^{12}$ & \rm $10^{-8}$ & \rm 9 & \rm $10^{-7}$\\  \hline 
\rm atoms & \rm molecule & \rm $10^{13}$ & \rm 2x$10^{-9}$ & \rm 7 & \rm $10^{-8}$\\  \hline 
\end{tabular}

\end{center}
\end{table}

\newpage

\begin{center}
{\bf Figures}
\end{center}

\noindent {\bf Figure 1:} One dimensional normal mode wave (solid line) on adiabatically
expanding closed loop (dotted line) representing cosmic metric. (After Peebles [14].) 

\begin{figure}[h]
\begin{center}
\includegraphics{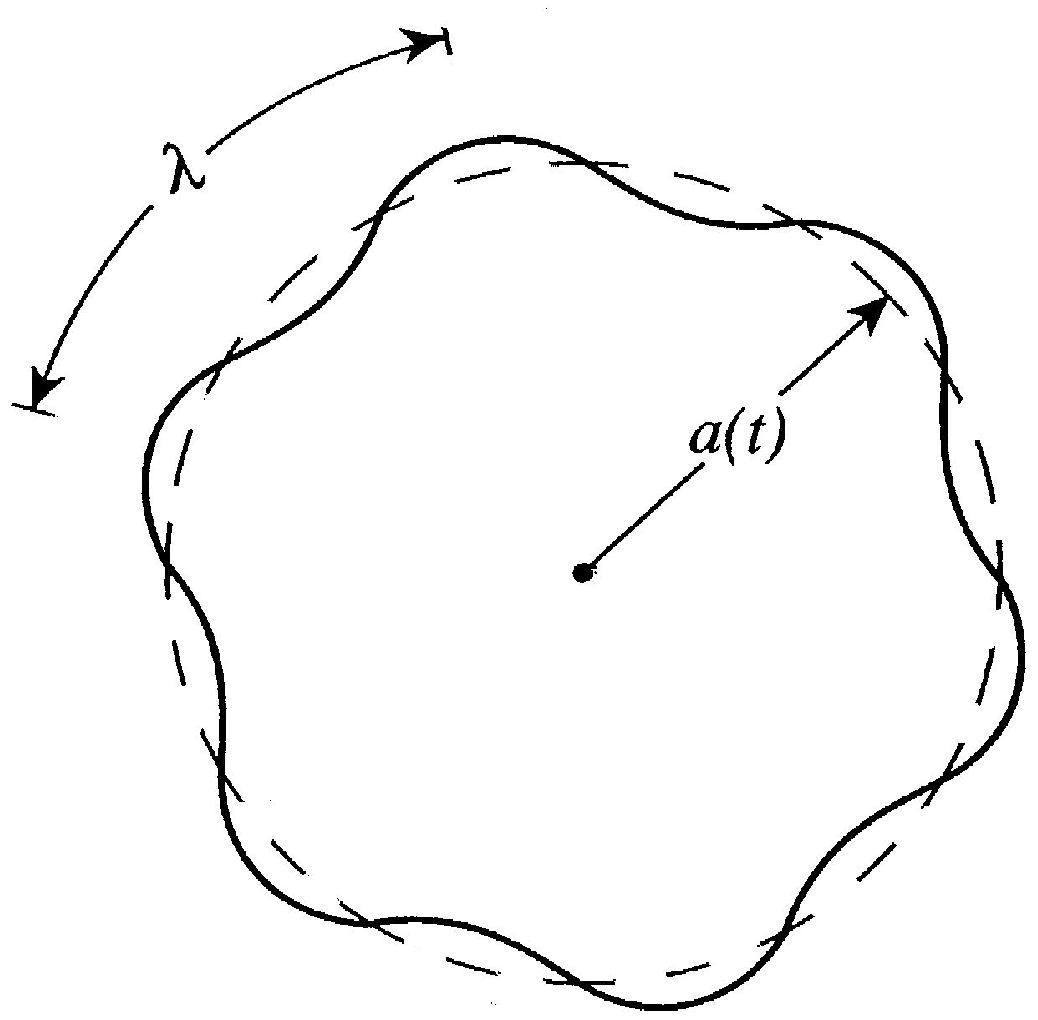}
\end{center}
\end{figure}

\end{document}